\documentclass[sn-mathphys]{sn-jnl}

\jyear{2021}%

\usepackage{adjustbox}  

\theoremstyle{thmstyleone}%
\theoremstyle{thmstyletwo}%

\theoremstyle{thmstylethree}%

\usepackage{tensor}

\newcommand{\TB}{\text{B}}
\newcommand{\tT}{\tilde{T}}
\newcommand{\tm}{\tilde{m}}

\newcommand{\tc}{\tilde{c}}

\newcommand{\tG}{\tilde{G}}

\newcommand{\te}{\tilde{e}}

\newcommand{\tlambda}{\tilde{\lambda}}

\newcommand{\tk}{\tilde{k}}

\newcommand{\tnu}{\tilde{\nu}}
\newcommand{\tmu}{\tilde{\mu}}

\newcommand{\tepsilon}{\tilde{\epsilon}}

\newcommand{\thbar}{\tilde{\hbar}}

\begin{document}

\title[Article Title]{A viable varying speed of light model in the RW metric}


\author*[1]{\fnm{Seokcheon} \sur{Lee}}\email{skylee@skku.edu}



\affil*[1]{\orgdiv{Department of Physics}, \orgname{Institute of Basic Science, Sungkyunkwan University,}, \orgaddress{ \city{Suwon}, \postcode{16419},  \country{Korea}}}




\abstract{The Robertson-Walker (RW) metric allows us to apply general relativity to model the behavior of the Universe as a whole (\textit{i.e.}, cosmology). We can properly interpret various cosmological observations, like the cosmological redshift, the Hubble parameter, geometrical distances, and so on, if we identify fundamental observers with individual galaxies. That is to say that the interpretation of observations of modern cosmology relies on the RW metric. The RW model satisfies the cosmological principle in which the 3-space always remains isotropic and homogeneous. One can derive the cosmological redshift relation from this condition. We show that it is still possible for us to obtain consistent results in a specific time-varying speed-of-light model without spoiling the success of the standard model.  The validity of this model needs to be determined by observations. }




\maketitle

\section{Introduction}\label{sec1}

Cosmology has been one of the most successful and relevant applications of Einstein's General Theory of Relativity (GR) to model the behavior of the Universe as a whole.  In other words,  cosmology as science can exist.  This fact becomes possible by adopting assumptions consistent with our observing Universe.  From the constancy of the temperature of cosmic microwave background (CMB) in a different direction in the sky,  we have good evidence that the Universe is  {\it isotropic} on the very largest scales \cite{Hinshaw:2013,Planck:2018nkj}.  If the Universe has no preferred center,  then the isotropy also implies the {\it homogeneity} consistent with the observation that the matter distribution looks uniform on scales of more than 100 million light-years \cite{Guzzo:2018xbe,DES:2020sjz}.  From these observations,  one reaches the {\it Cosmological Principle} (CP) that states that the Universe looks the same from all positions in space {\it at a particular time} and that all directions at any point are equivalent.  Thus,  it is possible to adopt the standard form of the Robertson-Walker (RW) metric in cosmology for the cosmic time $t$ \cite{Robertson:1929,Robertson:1933,Walker:1935,Walker:1937}.

The Lorentz transformation (LT) between two Galilean frames (GFs) derives consequences of special relativity (SR). In the general theory of relativity (GR), the inertial frame (IF) means a freely falling one. One can establish a Lorentz-invariant spacetime interval from the coordinate differences between two events. This spacetime interval is light-like ({\it i.e.}, null) if it equals zero. One can interpret this as signals moving at the speed of light connecting events in Minkowski spacetime separated by the null interval \cite{Morin07}. However, in GR, it is impossible to define a global time owing to the absence of a universal IF. Nevertheless, one can define a global time for the Universe when a set of requirements is satisfied and a metric embodying the cosmological principle meets these requirements. Then, one can define a global time by a foliation of spacetime as a sequence of non-intersecting spacelike 3D surfaces \cite{Weyl:1923,Islam02,Narlikar02,Hobson06,Gron07,Ryder09,CB15,Roos15,Guidry19,Ferrari21}. 

SR contains only one parameter, $c$, the speed of light in a vacuum.  We have shown that the universal Lorentz covariance, or, equivalently, the single postulate of Minkowski spacetime is good enough to satisfy the SR \cite{Das93,Schutz97,Lee:2020zts}. Thus, it is possible to make the Lorentz invariant (LI) varying speed of light (VSL) model as long as $c$ is locally constant and changes at cosmological scales. To avoid trivial rescaling of units, one must test the simultaneous variation of $c$ and Newton's gravitational constant $G$ because $c$ and $G$ enter as the combination $G/c^4$ in the Einstein action \cite{Barrow:1998eh}.  

All galaxies are assumed to lie on a hypersurface so that the surface of simultaneity for the local Lorentz frame (LF) of each galaxy coincides locally with the hypersurface. One can conclude that a global cosmic time for all galaxies on the hypersurface provided that space is homogeneous and isotropic (implying that the spatial curvature is constant). Cosmic time is the one measured by a comoving observer who observes the universe expanding uniformly around her. Light travels through the expanding space. The observation that sufficiently distant light sources show cosmological redshift corresponding to their distance from us follows the so-called Hubble’s law. We can induce the cosmological redshift relation from the cosmological principle (CP).

As a possible way to explain problematic observational results based on GR, the possibility of various VSLs has sometimes been invoked.  Einstein claimed that a shorter wavelength $\lambda$ leads to a lower speed of light using $c = \nu \lambda$ with the constant frequency $\nu$. He assumed that a gravitational field makes the clock run slower by $\nu_1 = \nu_2 (1 + GM/rc^2)$ \cite{Einstein:1911}. Dicke proposed that the wavelength and the frequency vary by defining a refractive index $n \equiv c/c_0 = 1 + 2GM/rc^2$ \cite{Dicke:1957}. He also considered a cosmology with an alternative description to the cosmological redshift by using a decreasing $c$ in time.  The assumption of these pioneer works is that the time dilation is due to the local gravity.  There have been cosmology-based VSL models to explain the horizon problem of the Big Bang model and provide an alternative method to cosmic inflation \cite{Barrow:1998eh,Petit:1988,Petit:1988-2,Petit:1989,Midy:1989,Moffat:1992ud,Petit:1995ass,Albrecht:1998ir,Barrow:1998he,Clayton:1998hv,Barrow:1999jq,Clayton:1999zs,Brandenberger:1999bi,Bassett:2000wj,Gopakumar:2000kp,Magueijo:2000zt,Magueijo:2000au,Magueijo:2003gj,Magueijo:2007gf,Petit:2008eb,Roshan:2009yb,Sanejouand:2009,Nassif:2012dr,Moffat:2014poa,Ravanpak:2017kdg,Costa:2017abc,Nassif:2018pdu}. A VSL model which proposed the change of the speed of light only without allowing the variations of other physical constants is called minimal VSL (mVSL).  Petit argued that if $c$ varies as a function of cosmic time, then one should include the joint variations of all related physical constants. These variations should be based on the consistency of all physical equations, and measurements of these constants remain consistent with physics laws during the evolution of the Universe. From this consideration, one might be able to obtain a universal gauge relationship and the temporal variation of the parameters that are regarded as constants \cite{Petit:1995ass,Petit:2008eb}
\begin{align}
G &= G_0 a^{-1} \,, \quad m = m_{0} a  \,, \quad c = c_0 a^{\frac{1}{2}}  \,, \quad h = h_0 a^{\frac{3}{2}} \,, \quad e = e_{0} a^{\frac{1}{2}}  \,, \quad \mu = \mu_0 a \label{Petitconst} \,.
\end{align}
We should emphasize those cosmology-based VSL models are different from those based on local gravity such as \cite{Einstein:1911,Dicke:1957}.  In these models (based on the RW metric), the cosmic time dilation is due to both the expansion of the Universe and the difference in the values of the local speed of light \cite{Barrow:1998he,Lee:2023rqv}. 

In Sec. ~\ref{sec:FLRW}, we review how to derive the cosmological redshift in the RW metric.  We extend this idea into the specific time-varying speed of light model (\textit{i.e.},  meVSL \cite{Lee:2020zts}) in section~\ref{sec:VSLz}. In section~\ref{sec:reason}, we explain how we can obtain consistent results in this model. We summarize the consequences of meVSL in the FLRW universe to keep the cosmological principle in section~\ref{sec:Cons}. We conclude in Sec. ~\ref{sec:Conc}. 

\section{Review of Cosmological redshift}
\label{sec:FLRW}

The RW metric is 
\begin{align}
ds^2 = -(d X^0)^2 +  a^2(t) \left( \frac{dr^2}{1-kr^2} + r^2 \left( d \theta^2 + \sin^2 \theta d \phi^2 \right) \right) \equiv -(d X^0)^2 +  a^2(t) dl_{3\textrm{D}}^2 \label{RW} \,,
\end{align}
where $X^0 = ct$.  In this metric, the light signal propagates along the null geodesic $ds^2 = 0$, and from metric \eqref{RW}, we obtain outgoing light signals
\begin{align}
dl_{3\textrm{D}}(r\,, \theta\,, \phi) = \frac{dX^0}{a(t)} \label{dl3D} \,.
\end{align}
The spatial infinitesimal line element $dl_{3\textrm{D}}$ is a function of the comoving coordinates ($\sigma\,,\theta\,,\phi$) only and thus should be the same value at any given time.  From this fact, one traditionally (\textit{i.e.}, by assuming a constant speed of light) obtains the cosmological redshift relation 
\begin{align}
\frac{dX^0(t_1)}{a(t_1)} = \frac{dX^0(t_2)}{a(t_2)}\quad \Rightarrow \quad \frac{dt_1}{a_1} = \frac{dt_2}{a_2} \quad \Rightarrow \quad \lambda_1 = c dt_1 \equiv c \nu_1^{-1} = \frac{a_1}{a_2} \lambda_2 \label{z} \,,
\end{align}
where $d t_i$ is the time interval of successive crests of light at $t_i$ (\textit{i.e.},  the inverse of the frequency $\nu_i$ at $t_i$) \cite{Weinberg:2008}.

\section{Cosmological redshift including VSL}
\label{sec:VSLz}

The standard cosmology has been very successful, and any viable new model should maintain results from the standard one. Now, we rederive the cosmological redshift relation in the meVSL model where we include the possibility of the varying speed of light at cosmological scales. Traditionally, we obtain the cosmological redshift under the assumption that the speed of light is constant at these scales.  However, the Lorentz invariance (LI) is a local symmetry that is only meaningful at each spacetime point (event), but GR is valid at cosmological scales. Therefore, the quibble about whether SR is generally adaptable at cosmological distances and time scales should be determined by observations \cite{Roos15}.  In this case, we can rewrite Eq.~\eqref{z} as
\begin{align}
\frac{dX^0(t_1)}{a(t_1)}  = \frac{dX^0(t_2)}{a(t_2)} \quad \Rightarrow \quad \frac{\tc_1 dt_1}{a_1} = \frac{\tc_2 dt_2}{a_2} \quad \Rightarrow \quad \lambda_1 = \tc_1 dt_1 = \frac{a_1}{a_2} \lambda_2 \label{zVSL} \,,
\end{align}
where $\tc_i \equiv \tc (t_i)$, $a_i \equiv a(t_i)$, and 
\begin{align}
dX^{0} = d \left(c t \right) = \left(\frac{d \ln c}{d \ln t} + 1 \right) c dt \equiv \tc dt \quad \textrm{and} \quad \delta c \equiv \frac{\tc}{c} = \left(\frac{d \ln c}{d \ln t} + 1 \right) \label{dX0} \,,
\end{align}
Thus, the cosmological redshift relation still holds even if we allow the speed of light can vary as a function of cosmic time.  These results have been derived before in a so-called meVSL model to satisfy Einstein field equations \cite{Lee:2020zts}.

\section{Reason for this possibility}
\label{sec:reason}

One measures the spectrum of light coming from a distant source.  One can determine the redshift by searching for features in the spectra like absorption lines, emission lines, or other variations in light intensity.  Cosmological redshift due to the expansion of the Universe may be characterized by the relative difference between the observed and emitted wavelengths of an object \cite{Weinberg:2008}.  The wavelength of the light at a given cosmic epoch $t_i$ is $\lambda_i$. Traditionally,  we make two assumptions. One is the clocks run at a different rate at different epochs $d t_i \neq d t_j$ when $i \neq j$, and the other is that the speed of light is constant during the entire history of the Universe.  But it is not a necessary condition to keep the LI because it is only true for the local inertial observer. There have been several projects to measure cosmological time dilation. Direct observation of the time dilation measures the decay time of distance supernova (SN) light curves and spectra \cite{Leibundgut:1996qm,SupernovaSearchTeam:1997gem,Foley:2005qu,Blondin:2007ua,Blondin:2008mz}. Another method is measuring time dilation by searching the stretching of peak-to-peak timescales of gamma-ray bursters (GRBs) \cite{Norris:1993hda,Wijers:1994qf,Band:1994ee,Meszaros:1995gj,Lee:1996zu,Chang:2001fy,Crawford:2009be,Zhang:2013yna,Singh:2021jgr}. There has been a search for the time dilation effect in the light curves of quasars (QSOs) located at cosmological distances \cite{Hawkins:2001be,Dai:2012wp}.  So far, it seems fair to say that no convincing detection has been made for cosmic time dilation with the conflict between different measurements. Also, there is no mechanism to determine $dt_i$ from the RW model.  Therefore, it might still be meaningful to investigate the possibility of varying speed of light in this observation as long as its results are consistent with the standard model ones.

\section{Consequences}
\label{sec:Cons}

If $c$ is not a constant in cosmic time, then one needs to rederive the Einstein field equations again by including this VSL effect.  Previously, we have shown one viable model \cite{Lee:2020zts}.  Metric and four-position are given by 
\begin{align}
g_{\mu\nu} = \textrm{diag} \left( -1\,,  \frac{a^2}{1-kr^2}\,, a^2 r^2\,, a^2 r^2 \sin^2 \theta \right) \quad , \quad x^{\mu} = \left( c[a]t \,, x \,, y \,, z  \right) \label{gmunuxmu} \,,
\end{align}
where we will denote the 3D spatial metric as $\gamma_{ij}$. Christoffel symbols are obtained from their definition
\begin{align}
&\Gamma^{\mu}_{\nu\lambda} \equiv \frac{1}{2} g^{\mu\alpha} \left( g_{\alpha\nu,\lambda} + g_{\alpha\lambda,\nu} - g_{\nu\lambda,\alpha} \right) \label{GammaApp} \, \\
&\Gamma^{0}_{ij} \equiv \frac{1}{2} g^{00} \left( g_{0j,i} + g_{i0,j} - g_{ij,0} \right) = \frac{1}{2} (-1) \left(- \frac{d a^2}{d x^0}  \gamma_{ij}\right) = \frac{a \dot{a}}{\tc} \gamma_{ij} \label{Gamma0ij} \\ &\Gamma^{i}_{0j} = \frac{1}{\tc}  \frac{\dot{a}}{a} \delta^i_j \quad , \quad \Gamma^{i}_{jk} = ^{s}\Gamma^{i}_{jk}  \label{GammacompApp} \,,
\end{align}
where $^{s}\Gamma^{i}_{jk}$ denote the Christoffel symbols for the spatial metric $\gamma_{ij}$ and we use
\begin{align}
dx^0 = d \left(c[a] t] \right) = \left( \frac{d \ln c}{d \ln a} \frac{d \ln a}{dt} t + 1 \right) c dt =  \left( \frac{d \ln c}{d \ln a}  H t + 1 \right) c dt 
\equiv \tc dt \label{dx0} \,. \end{align}
One obtains Riemann curvature tensors, Ricci curvature tensors, and a Ricci scalar.
\begin{align}
&\tensor{R}{^\alpha_\beta_\mu_\nu} = \Gamma^{\alpha}_{\beta\nu,\mu} - \Gamma^{\alpha}_{\beta\mu,\nu} + \Gamma^{\alpha}_{\lambda\mu} \Gamma^{\lambda}_{\beta\nu} - \Gamma^{\alpha}_{\lambda\nu} \Gamma^{\lambda}_{\beta\mu} \label{RabmuApp} \,, \\
&\tensor{R}{^0_i_0_j} = \frac{\gamma_{ij}}{\tc^2} \left( a \ddot{a} - \dot{a}^2 \frac{d \ln \tc}{d \ln a} \right) \quad , \quad \tensor{R}{^i_0_0_j} = \frac{\delta^{i}_{j}}{\tc^2} \left( \frac{\ddot{a}}{a} - \frac{\dot{a}^2}{a^2} \frac{d \ln \tc}{ d \ln a}  \right) \,, \label{R0i0jApp} \\
&\tensor{R}{^i_j_k_m} = \frac{\dot{a}^2}{\tc^2} \left( \delta^{i}_{k} \gamma_{jm} - \delta^i_m \gamma_{jk} \right) + \tensor[^s]{R}{^i_j_k_m} \quad , \quad \tensor[^s]{R}{^i_j_k_m} = k \left( \delta^i_k \gamma_{jm} - \delta^i_m \gamma_{jk} \right) \,. \label{RijkmApp} \\
&R_{\mu\nu} = \Gamma^{\lambda}_{\mu\nu,\lambda} - \Gamma^{\lambda}_{\mu\lambda,\nu} + \Gamma^{\lambda}_{\mu\nu} \Gamma^{\sigma}_{\lambda\sigma} - \Gamma^{\sigma}_{\mu\lambda} \Gamma^{\lambda}_{\nu \sigma} \label{RmnApp} \,, \\
&R_{00} = -\frac{3}{\tc^2} \left( \frac{\ddot{a}}{a} - \frac{\dot{a}^2}{a^2} \frac{d \ln \tc}{ d \ln a}  \right) \quad , \quad
R_{ij} = \frac{\gamma_{ij}}{\tc^2} a^2 \left( 2 \frac{\dot{a}^2}{a^2} + \frac{\ddot{a}}{a} + 2 k \frac{\tc^2}{a^2} - \frac{\dot{a}^2}{a^2} \frac{d \ln \tc}{ d \ln a}  \right) \label{RijApp} \,, \\
&R = \frac{6}{\tc^2} \left( \frac{\ddot{a}}{a} + \frac{\dot{a}^2}{a^2} + k \frac{\tc^2}{a^2} - \frac{\dot{a}^2}{a^2} \frac{d \ln \tc}{ d \ln a}  \right) \label{RmpApp} \,.
\end{align} 
It causes the modification of the usual Friedmann equations in the VSL model shown in the previous work \cite{Lee:2020zts}. All of this argument relies on an RW metric and requires the homogeneity and isotropy of the 3D space. The adiabatic expansion of the Universe is a prerequisite to keep the CP. As $c$ evolves as a function of cosmic time, other physical constants and quantities also need to do as a function of $t$ to satisfy the CP \cite{Lee:2212}. In the meVSL model, one obtains the same result as a constant $c$ calculation for the comoving distance $l_{3\textrm{D}}$ by integrating $dl_{3\textrm{D}}$ for the cosmological redshift $z$ 

\begin{align}
\Delta s^2 = 0 \Rightarrow d l_{3\textrm{D}} = \frac{\tilde{c} dt}{a} = \frac{\tilde{c} da}{a^2 H(a)} = -\frac{c_0 dz}{H^{(\text{GR})}(z)} 
\end{align}
where $H(a) = H^{(\text{GR})} a^{b/4}$ when $\tilde{c} = c_0 a^{b/4}$.  $H^{(\text{GR})}$ denotes the Hubble parameter when $c$ is a constant. Thus,  the Hubble radius $c/H$ is the same as that of GR for the meVSL model and does not solve the horizon problem of the standard big bang model. However, the Hubble parameter is modified from that of GR by the extra factor $a^{b/4}$. It might solve the $H$-tension problem. In the meVSL model, we consider local thermodynamics, energy conservation, and other local physics. These considerations induce the time evolutions of other physical constants and quantities shown in table~\ref{tab:table-1}.

\begin{table}[htbp]
\caption{Summary for cosmological evolutions of physical constants and quantities of the meVSL model. These relations satisfy all known local physics laws, including special relativity, thermodynamics, and electromagnetic force \cite{Lee:2020zts,Lee:2212}. One should remind that the wavelength, frequency, and temperature of the photon (dimensionful physical quantities) evolve as cosmic time even in the standard model.}
\label{tab:table-1}
\begin{adjustbox}{width=\columnwidth,center}
\begin{tabular}{|c||c|c|c|}
	\hline
	local physics laws & Special Relativity & Electromagnetism & Thermodynamics \\
	\hline \hline
	quantities & $\tm = \tm_0 a^{-b/2}$ & $\te = \te_0 a^{-b/4}\,, \tlambda = \tlambda_0 a \,, \tnu = \tnu_0 a^{-1+b/4}$ & $\tT = \tT_0 a^{-1}$ \\
	\hline
	constants & $\tc = \tc_0 a^{b/4} \,, \tG = \tG_0 a^{b}$ & $\tepsilon = \tepsilon_0 a^{-b/4} \,, \tmu = \tmu_0 a^{-b/4} \,, \tc = \tc_0 a^{b/4}$ & $\tk_{\TB 0} \,, \thbar = \thbar_0 a^{-b/4}$ \\
	\hline
	energies & $\tm \tc^2 = \tm_0 \tc_0^2$ & $\tilde{h} \tnu = \tilde{h}_0 \tnu_0 a^{-1}$ & $\tk_{\TB} \tT = \tk_{\TB 0} \tT_0 a^{-1}$ \\
	\hline
\end{tabular}
\end{adjustbox}
\end{table}

\section{Conclusion}\label{sec:Conc}

We interpret most current cosmological observations based on the $\Lambda$CDM cosmological model. It uses the RW metric, and it is valuable for us to clarify any possibility of the extension of this model. We show that the meVSL model provides consistent results for various measurements even if we allow the speed of light to change in the expanding Universe. However, other physical constants, including the Planck constant, should also evolve as a function of cosmic time to be a viable model. The consequences of modifications of cosmological observations compared to the standard model would determine the validity of the meVSL model.

\bmhead{Acknowledgments}

SL is supported by Basic Science Research Program through the National Research Foundation of Korea (NRF) funded by the Ministry of Science, ICT, and Future Planning (Grant No. NRF-2017R1A2B4011168 and No. NRF-2019R1A6A1A10073079). SL appreciates professor C. Ahn, K. Ahn, S. Appleby, G. F. R. Ellis, B.S.~Kyae, K. W.Ng, K. A. Olive, W. I.  Park, and S.Y. Tsai  for their useful discussions and comments. 

\section*{Availability of data and materials}

Data sharing is not applicable to this article as no new data were created or analyzed in this study.









\end{document}